\documentclass[conference]{IEEEtran}
\IEEEoverridecommandlockouts
\usepackage{cite}
\usepackage{amsmath,amssymb,amsfonts}
\usepackage{algorithmic}
\usepackage{bbm}
\usepackage{gensymb}
\usepackage{graphicx}
\usepackage{textcomp}
\usepackage{xcolor}
\usepackage{subfigure}

\usepackage{array}
\usepackage{pifont}
\newcommand{\cmark}{\textcolor{green!80!black}{\ding{51}}}
\newcommand{\xmark}{\textcolor{red}{\ding{55}}}

\def\BibTeX{{\rm B\kern-.05em{\sc i\kern-.025em b}\kern-.08em
    T\kern-.1667em\lower.7ex\hbox{E}\kern-.125emX}}
\begin{document}

\title{Managing Charging Induced Grid Stress and Battery Degradation in Electric Taxi Fleets
\thanks{The research is funded by the Indonesian Endowment Fund for Education (LPDP) on behalf of the Ministry of Education, Culture, Research, and Technology of the Republic of Indonesia and managed by Institut Teknologi Bandung under the INSPIRASI Program (Contract No. 2963/E3/AL.04/2024 and 314/IT1.B07/KS.00/2024).}
}

% \author{\IEEEauthorblockN{Michael Yuhas}
% \IEEEauthorblockA{\textit{Energy Research Institute} \\
% \textit{Nanyang Technological University}\\
% Singapore \\
% michael.yuhas@ntu.edu.sg}
% \and
% \IEEEauthorblockN{Rajesh K. Ahir}
% \IEEEauthorblockA{\textit{Energy Research Institute} \\
% \textit{Nanyang Technological University}\\
% Singapore \\
% rajeshkumar.ahir@ntu.edu.sg}
% \and
% \IEEEauthorblockN{Laksamana Vixell Tanjaya Hartono}
% \IEEEauthorblockA{\textit{Electrical Engineering and
% Informatics} \\
% \textit{Bandung Institute of Technology}\\
% Bandung, Indonesia \\
% n2409706a@e.ntu.edu.sg}
% \and
% \IEEEauthorblockN{Muhammad Dzaki Dwi Putranto}
% \IEEEauthorblockA{\textit{Electrical Engineering and Informatics} \\
% \textit{Bandung Institute of Technology}\\
% Bandung, Indonesia \\
% dzaki.dwi@itb.ac.id}
% \and
% \IEEEauthorblockN{Arvind Easwaran}
% \IEEEauthorblockA{\textit{College of Computing and Data Science} \\
% \textit{Nanyang Technological University}\\
% Singapore \\
% arvinde@ntu.edu.sg}
% \and
% \IEEEauthorblockN{Suhono Harso Supangkat}
% \IEEEauthorblockA{\textit{Electrical Engineering and Informatics} \\
% \textit{Bandung Institute of Technology}\\
% Bandung, Indonesia \\
% suhono@itb.ac.id}
% }
\author{
    \IEEEauthorblockN{
        Michael Yuhas\IEEEauthorrefmark{1},
        Rajesh K. Ahir\IEEEauthorrefmark{1},
        Laksamana Vixell Tanjaya Hartono\IEEEauthorrefmark{2},
        Muhammad Dzaki Dwi Putranto\IEEEauthorrefmark{2},\\
        Arvind Easwaran\IEEEauthorrefmark{3}\IEEEauthorrefmark{1}
        Suhono Harso Supangkat\IEEEauthorrefmark{2}
    }
    \IEEEauthorblockA{\IEEEauthorrefmark{1}\textit{Energy Research Institute, Nanyang Technological University}, Singapore
    \\\{michael.yuhas, rajeshkumar.ahir, arvinde\}@ntu.edu.sg}
    \IEEEauthorblockA{\IEEEauthorrefmark{2}\textit{School of Electrical Engineering and Informatics, Institut Teknologi Bandung},
Bandung, Indonesia,\\ \{23524067, dzaki.dwi, suhono\}@itb.ac.id}

    \IEEEauthorblockA{\IEEEauthorrefmark{3}\textit{College of Computing and Data Science, Nanyang Technological University}, Singapore}
}

\maketitle

\begin{abstract}
Operating fleets of electric vehicles (EVs) introduces several challenges, some of which are borne by the fleet operator, and some of which are borne by the power grid.  To maximize short-term profit a fleet operator could always charge EVs at the maximum rate to ensure vehicles are ready to service ride demand. However, due to the stochastic nature of electricity demand, charging EVs at their maximum rate may potentially increase the grid stress and lead to overall instability.  Furthermore, high-rate charging of EVs can accelerate battery degradation, thereby reducing the service lifespan of the fleet. This study aims to reconcile the conflicting incentives of fleet longevity, short-term profitability, and grid stability by simulating a taxi fleet throughout its lifespan in relation to its charging policies and service conditions. We develop an EV fleet simulator to evaluate the battery degradation due to unpredictable charging and ride demand. Consequently, the impact on the power grid through the charging infrastructure is assessed due to these activities. This simulation utilizes publicly accessible real-world travel data from the NYC taxi dataset.  We compare a baseline 80-20 fleet charging policy with a reinforcement learning-based policy designed to prolong the fleet’s service life and alleviate grid stress.  We monitor grid stress, battery degradation, and profitability over five years and find that our learned policy outperforms the baseline. This simulator enables fleet operators to assess the impact of different charging policies on these indicators to make informed decisions in the future. 
\end{abstract}


\section{Introduction}
Electric vehicles (EVs) comprise a growing portion of taxi fleets, particularly in developing economies where charging infrastructure and grid development may not match the pace of deployment. For instance, in Indonesia, BlueBird, a major EV fleet taxi operator, plans to deploy more than 1000 EVs in Jakarta by 2025 to replace the fossil fuel-based taxis~\cite{rusli2024}. Although the country as a whole faces an oversupply of electricity, localized constraints such as last-mile charging infrastructure remain a significant barrier~\cite{syaharani2024}. At the same time, the daily demand for rides far exceeds the capacity of the current EV taxi fleets~\cite{sari2025}, while the absence of dynamic pricing in the Indonesian grid~\cite{kapanlagi} means that the challenge for fleet operators is not to minimize charging costs. Instead, the central issue lies in scheduling charging operations so that taxis are available to meet demand in both time and location, without overloading local grid capacity at charging points or accelerating EV battery degradation.

To address this challenge, we consider three holistic indicators: grid stress, battery degradation, and fleet revenue (profitability).  We consider grid stress as the instantaneous power demand to charge the fleet at any point in time.  Battery degradation is the reduction in a batteries peak capacity over time, also referred to as state of health (SoH)~\cite{preger2020}. We consider fleet revenue only from servicing ride demand and ignore other operational costs such as the price of electricity (constant in the Indonesian context) and fixed costs.  These three indicators have a complex relationship where maximizing one may adversely affect the others.
For example, while charging the EVs at a faster rate and to maximum capacity helps maximize the short-term profitability of the fleet, it expedites the battery deterioration process as well. Over time, this deterioration reduces the fleet's ability to meet demand, leaving portions of demand unserviceable, leading to a reduction in profitability.
Charging rates affect availability -- the ability of fleet vehicles to service demand when it arrives~\cite{abdi2020} -- by affecting battery aging~\cite{khan2023}. However, charging rates are influenced in turn by localized grid conditions~\cite{yuan2022}, as well as the number of vehicles and their locations relative to charging stations at any given time~\cite{lee2021}.

\begin{figure}
    \centering
    \includegraphics[width=0.8\linewidth]{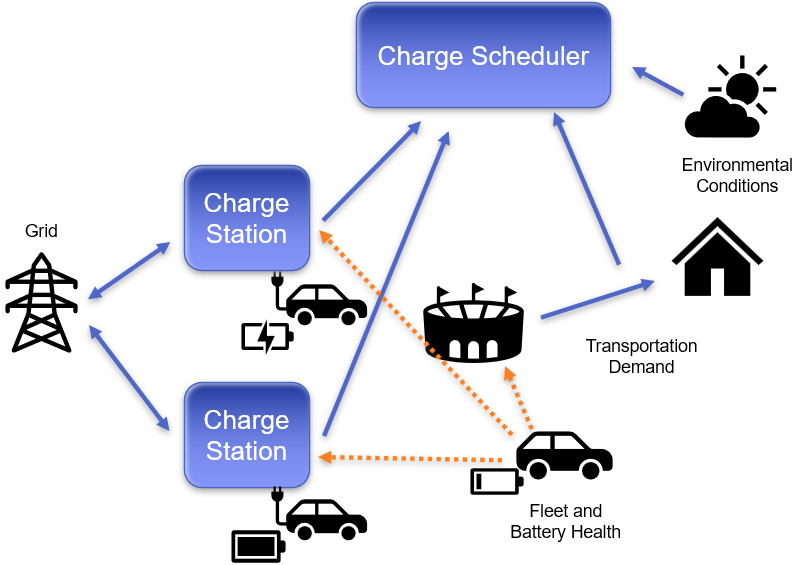}
    \caption{A charge scheduler must decide where, when, and at what rate vehicles charge based on grid conditions, charger availability, ride demand, fleet location and health, and environmental conditions.  Orange lines indicate the set of options the charge scheduler has for the selected vehicle.}
    \label{fig:intro}
\end{figure}

There is no closed form solution to model these three interacting indicators over the long horizon.
Fig.~\ref{fig:intro} illustrates the relationships between some of the external factors affecting fleet performance over time.
Environmental conditions impact the grid and charging infrastructure as well as the efficiency of the EVs themselves~\cite{ou2023}, and ride demand varies based on date, time of day, and weather conditions~\cite{lepage2021}.
A charge scheduler that tells EVs where, when, and how fast to charge needs to take all these factors into account in addition to the current location, state of charge (SoC), and SoH of each vehicle in the fleet.
The decisions made by the charge scheduler then go on to impact future fleet states, which impact future scheduling decisions.
Ultimately, this impacts performance with respect to the three aforementioned indicators.

To optimize these indicators in the face of such a complex relationship, we propose a simulation-based approach that will allow us to model different charge scheduling policies in the context of ride demand, long term battery degradation, and grid/infrastructure constraints.
Existing works have simulated battery aging~\cite{yao2020}, charge scheduling~\cite{lee2019}, and fleet dispatch~\cite{zhang2024} independently, but to our knowledge, this is the first work that combines all three into one simulator.
We then use this simulator to evaluate a widely used baseline charging policy based on the 80-20 charging rule~\cite{davies2019}. In addition, we use the simulator to train a simple reinforcement learning solution that outperforms the baseline in terms of grid stress, lifetime revenue, and battery degradation.
In summary, the novelty of this work is threefold: 
\begin{enumerate}
    \item We use long term profitability, grid stress, and battery health to assess EV fleet charge scheduling policies over a multi-year horizon.
    \item We develop a simulator that combines demand modelling, charging infrastructure, and battery health to provide the in-depth assessment.
    \item We use the simulator to examine a simple reinforcement learning-based charging policy that outperforms a widely used baseline.
\end{enumerate}

\section{Related Work}
Charge scheduling, battery degradation, and fleet dynamics have all been explored independently in previous literature and various tools and datasets exist that address these problems in isolation. 
We leverage some of these previous approaches to solve the more complex, integrated problem.
Table~\ref{tab:litreview} summarizes the capabilities of existing simulators and demonstrates the gap filled by this work.

\subsection{Charge Scheduling}
The infrastructure required to charge EVs is not only constrained by the number of ports available for simultaneous EV charging, but the total amount of power that can be drawn by a single charging station. Such a constraint could come from the physical size of the on-site transformer or due to grid status at a certain time. Lee et al. investigated such constraints by proposing an adaptive charge network simulator~\cite{lee2019}, which simulates a network of chargers with vehicles arriving randomly and then built upon this to minimize the makespan of a set of charging vehicles~\cite{lee2021}. However, the current simulator does not model the effect of charging decisions on future vehicle arrivals. For example, in a taxi fleet, terminating a vehicle's charging session at a certain time releases it to service demand, which when coupled with task arrivals, influences when the vehicle will return to charge. Additionally, several efforts have been made to consider the possibility of EVs supplying the grid during peak loading hours~\cite{mukherjee2015} or vehicle-to-vehicle charge scheduling within a station~\cite{dan2024}.

\subsection{Battery Degradation}
Battery degradation comes from two sources.  One is calendar aging, where a battery degrades over time as a function of its SoC~\cite{keil2016}. The other is thermal aging, which occurs every time a battery is charged or discharged~\cite{perez2017}. Many degradation models have been proposed for lithium-ion batteries~\cite{elmahallawy2022}, but we will focus on the recent multi-stage degradation model where the rate of degradation varies depending on the current state of health~\cite{wan2024}. Previous studies have developed simulators for long-term battery degradation in EVs. The authors have considered driving conditions and holistically modelled the effect of temperature, number of passengers, and road conditions on a vehicle's energy expenditure~\cite{ou2023}. Yao et al. also modelled battery degradation in a taxi fleet and considered operational costs such as charging and parking over the long term~\cite{yao2020}. However, this simulator overlooked the vehicle’s distance to ride demands and grid constraints beyond a charger's rated capacity.

\begin{table}
  \centering
  \caption{Capability comparison among simulators. Vehicle ops. refers to costs incurred servicing demand, while vehicle routing refers to the ability to evaluate different paths in simulation.}
  \label{tab:litreview}
  \begin{tabular}{p{0.24\linewidth}|p{0.125\linewidth}|p{0.125\linewidth}|p{0.125\linewidth}|p{0.125\linewidth}}
    \hline
    \textbf{Simulator}&\textbf{Charging Rate}&\textbf{Battery Deg.}
    &\textbf{Vehicle Ops.}&\textbf{Vehicle Routing}\\
    \hline
    ACN-Sim~\cite{lee2019}& \cmark & \xmark & \xmark & \xmark \\
    %\hline
    BREVO~\cite{ou2023}& \xmark & \cmark & \cmark & \xmark \\
    %\hline
    Yao et al.~\cite{yao2020} & \cmark & \cmark & \cmark & \xmark \\
    %\hline
    TaxiSim~\cite{cheng2011} & \xmark & \xmark & \xmark & \cmark \\
    %\hline
    Bischoff et al.~\cite{bischoff2014} & \cmark & \xmark & \xmark & \cmark \\
    %\hline
    Zhang et al.~\cite{zhang2024} & \xmark & \xmark & \cmark & \cmark \\
    %\hline
    FleetRL~\cite{cording2024} & \cmark & \cmark & \cmark & \xmark \\
    %\hline
    Ours&\cmark &\cmark & \cmark & \cmark \\
  \hline
\end{tabular}
\end{table}

\subsection{Fleet Simulation}
Unlike~\cite{lee2019} and~\cite{yao2020}, this study considers where and how long vehicles must travel before and after charging. TaxiSim models fleet vehicles as agents with a finite number of states and user-defined ride demand service policies but does not consider charging or grid conditions~\cite{cheng2011}. TaxiSim requires real-world demand data to simulate demand requests (every time a customer hails a ride). This can be provided by real-world datasets such as the New York City Taxi Dataset~\cite{nyc-dataset} and the City of Chicago Taxi Rides Dataset~\cite{chicago-dataset}. Bischoff et al. developed a simulator for EV taxi fleets, but it did not consider long-term degradation or dynamic grid conditions~\cite{bischoff2014}. Recently, Zhang et al. also developed an EV taxi fleet simulator with a focus on mapping taxis to rides and chargers~\cite{zhang2024}, but their work overlooked battery health or the ability to schedule different charging rates.  Cording et al. developed FleetRL to train reinforcement learning models for EV charging optimization, but this simulator does not consider the path vehicles travel after they are released from the charger to serve demand~\cite{cording2024}.

\section{Long Horizon Modeling of Electric Taxi Fleets with a Simulator}
\begin{figure}
    \centering
    \includegraphics[width=1\linewidth]{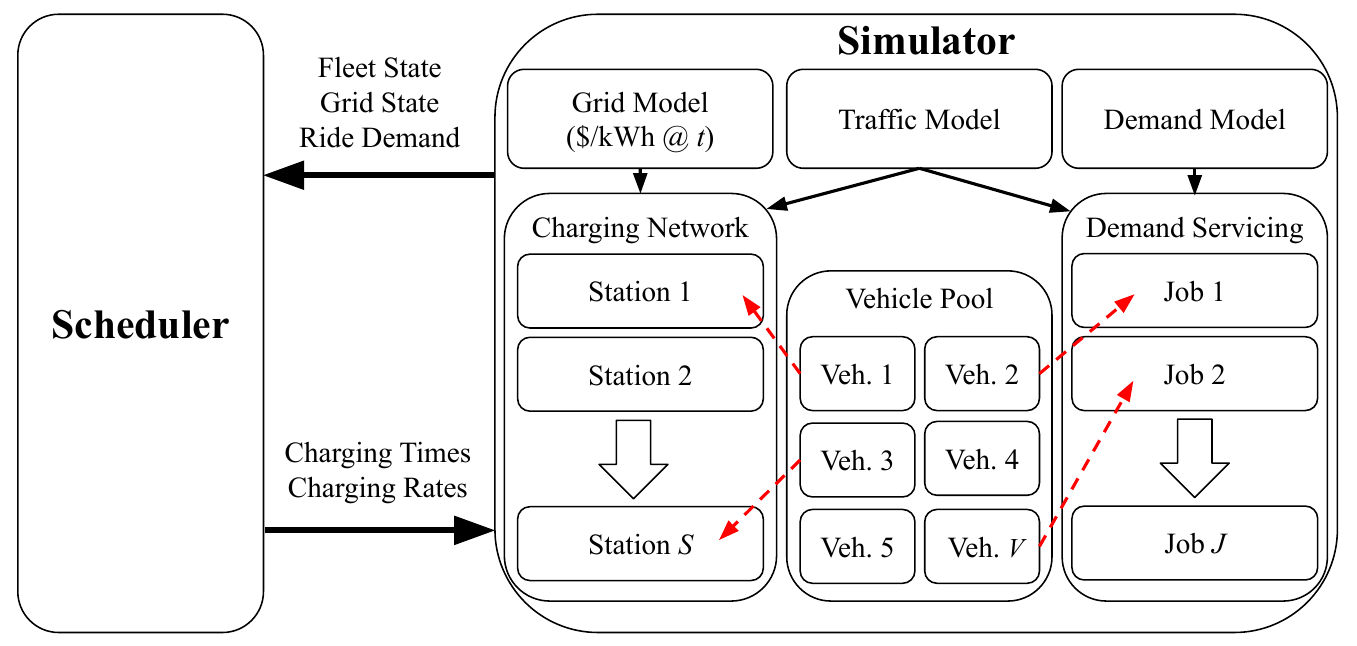}
    \caption{High level block diagram of our simulator.  Vehicles are persistent software objects that can be assigned to charge, service demand, or idle by the scheduler.}
    \label{fig:block_diagram}
\end{figure}

This section describes how the proposed simulator models electric taxi fleets over the long horizon and how to evaluate performance across the fleet's life. Table~\ref{tab:notation} provides a list of the notation used. Fig.~\ref{fig:block_diagram} shows a high-level block diagram of the proposed simulator. The vehicle pool is a static list of persistent vehicle objects, each of which maintains its own state. A demand model generates jobs (ride requests) which must be assigned to a vehicle and cost a certain amount of time and energy to complete. The completion time of a job is dependent on traffic conditions, which also influence the time and energy required for a vehicle to reach a charging station. Vehicles can also be assigned to chargers to replenish SoC, with the maximum charging rate dependent on the grid state. The simulator recalculates the internal state of each of its components at discrete points in time. At each time $t$ the state is communicated to a scheduler, which returns a list of vehicle assignments to ride demand and chargers, as well as their charging rates. The scheduler must take into account the current grid state so that charging several vehicles at once does not overstress the power system.

\begin{table}
  \caption{Table of Simulator Notation}
  \label{tab:notation}
  \begin{tabular}{p{0.15\linewidth} p{0.75\linewidth}}
    \hline
    \textbf{Symbol}&\textbf{Notation}\\
    \hline
    $\mathcal{V}$ & Set of vehicles in a taxi fleet.\\
    $v_j/v_p$ & vehicle assigned to job $j$ or charge port $p$.\\
    $Q_v(t)$ & State of charge in vehicle $v$'s battery at time $t$.\\
    $\bar{Q}_{v}(t)$ & State of health of vehicle $v$'s battery at time $t$.\\
    $\Delta Q_v(t)$ & Loss in capacity of $v$'s battery on the interval $[t-1,t)$.\\
    $N_{cref}$ & Reference number of charge cycles to battery end-of-life.\\
    $\alpha, \beta, \psi$ & Health dependent battery parameters.\\
    $C_v(t)$ & C-rate used to charge or discharge $v$'s battery on $[t-1,t)$.\\
    $T_{ref}$ & Reference temperature ($25\degree C$).\\
    $T_v(t)$ & Temperature on vehicle $v$'s battery at time $t$.\\
    $\eta_v$ & Vehicle $v$'s efficiency in charge per kilometer.\\
    $\mathcal{J}$ & Set of jobs (ride-demand).\\
    $r_j$ & Arrival time of job $j$.\\
    $l_{pu,j}/l_{do,j}/l_s$ & Location of pick-up/drop-off for job $j$ or charge station $s$.\\
    $\sigma_j$ & state of job $j$ \\
    $\mathcal{S}$ & Set of charging stations.\\
    $\mathcal{P}$ & Set of ports comprising a charging station.\\
    $\bar{P}_s/\bar{P}_p$ & Maximum power supported by station $s$ or port $p$.\\
    $P_p(t)$ & Output power of port $p$ on the interval $[t-1,t)$.\\
    $\eta_s$ & Efficiency of charging station $s$.\\
  \hline
\end{tabular}
\end{table}

\subsection{Vehicle Pool}
We consider a fleet $\mathcal{V}$ with individual vehicles represented as $v\in\mathcal{V}$. All vehicles have a state which respects the transition rules in Fig.~\ref{fig:veh_sd}: only \textsc{idle} vehicles can be assigned to charge or service demand, and the travel times to chargers or pick-up locations are tracked in separate states.  If a vehicle does not have sufficient charge to complete its travel to its destination, it enters the \textsc{recovery} state.  This state indicates the vehicle cannot travel for the next 24 hours after which it is moved to its assigned depo in the \textsc{idle} state with full charge. At any time $t$, the SoC of a vehicle's battery is represented by $Q_v(t)$. Additionally, every vehicle's battery has a state of health represented by $\bar{Q}_v(t)$, i.e., the maximum amount of charge that the battery could possibly contain at $t$. Both $Q_v(t)$ and $\bar{Q}_v(t)$ are given in units of energy (kWh). Over the lifespan of the vehicle, $\bar{Q}_v(t)$ decreases due to calendar aging and thermal aging, however, since all vehicles in the fleet will undergo calendar aging at roughly the same rate, we only consider degradation due to thermal aging for now~\cite{wan2024}. 
Furthermore, empirical results show that when a battery's SoC is cycled continuously, thermal aging dominates calendar aging~\cite{liu2024}, which is the case in our experiments where vehicle idle times are small.
The multi-stage battery degradation model proposed by Wan et al.~\cite{wan2024} calculates a capacity loss over the battery's life on the interval $[0,t)$. We adapt this model to work with our tick-based simulator, by calculating a discrete capacity loss $\Delta Q_v(t)$ on every interval $[t-1, t)$. The expression for $\Delta Q_v(t)$ is given in eq.~\ref{eq:delQ} where $N_{cref}=513$ is the number of cycles to end-of-life under a reference cycle condition, $\alpha$ and $\beta$ are stage dependent parameters, $\psi$ is the Arrhenius rate constant, $C_v(t)$ is the C-rate~\cite{khan2023} used to charge or discharge a vehicle's battery on the interval $[t-1,t)$, $T_{ref}$ is the reference temperature ($25\degree C$), and $T_v(t)$ is the temperature of vehicle $v$'s battery at time $t$.
\begin{align*}
 \Delta Q_v(t) = \frac{1}{N_{cref}}
    \times \Big( 1-\frac{Q_v(t)}{\bar{Q}_v(t)}\Big)^{-1/\alpha}
    \times \Big( \frac{2C_v(t)}{\Delta t}\Big)^{-1/\beta}\\
    \times exp\Big( -\psi\big(\frac{1}{T_{v}(t)} - \frac{1}{T_{ref}}\big)\Big)
    \addtocounter{equation}{1}\tag{\theequation}
    \label{eq:delQ}
\end{align*}
The stage dependent parameters are listed in Table~\ref{tab:batparams} with SoH representing a percentage of the battery's initial capacity $\bar{Q}_v(t)/\bar{Q}_v(0)$~\cite{preger2020}. Thus, at any time $t$, the maximum charge vehicle $v$'s battery can store is given by eq.~\ref{qmax}.
\begin{equation}\label{qmax}
    \bar{Q}_{v}(t)=\bar{Q}_{v}(t-1)-\Delta \bar{Q}_v(t)
\end{equation}
Using the C-rate, we can also calculate the actual charge in a vehicle's battery at $t$ using eq.~\ref{eq:qact}.  Here, $\Delta t$ is the simulator tick length, which can be decreased to improve simulation accuracy or increased to improve modelling efficiency.
\begin{equation}\label{eq:qact}
    Q_v(t) = Q_v(t-1) + C_v(t)\Delta t
\end{equation}
C-rate during charging time intervals is determined by the charge schedule, while C-rate during service intervals is calculated by multiplying the distance covered by the vehicle by an efficiency factor $\eta_v$, which describes the amount of charge needed to cover a given distance for a specific vehicle type.
\begin{table}
\centering
  \caption{Multi-stage battery degradation parameters measured by Preger et al.~\cite{preger2020}.  SoH indicates the ratio $\bar{Q}_v(t)/\bar{Q}_v(0)$.}
  \label{tab:batparams}
  \begin{tabular}{c c c c c}
    \hline
    \textbf{Stage}&\textbf{SoH}&$\alpha$&$\beta$&$\psi$\\
    \hline
    $1$ &   $100\% - 93.3\%$ & 0.2171 & 24.2535 & -12.0051\\
    $2$ &   $93.3\% - 86.6\%$ & 0.2652 & 9.9653 & -29.0049\\
    $3$ &   $86.6\% - 0\%$ & 0.2611 & -15.1963 & -22.5247\\
    \hline
\end{tabular}
\end{table}  

\subsection{Traffic Model}
Calculating the precise distance and time a vehicle needs to travel between specific points for every journey would be intractable across a multi-year simulation.
To solve this problem, we divide the region in which a fleet operates into a discrete number of zones and precalculate the travel time and distance between each zone.
The traffic model can be thought of as a function $f(l_{start},l_{end}):\mathbb{N}\times\mathbb{N}\rightarrow \mathbb{R}_+^2$ where time and distance can be any positive real number.
This works well with existing ride demand datasets that divide urban areas into zones, even when precise pick-up and drop-off coordinates are provided~\cite{nyc-dataset,chicago-dataset}.
In this work, we iterate through the New York Taxi Dataset and use a kernel density estimate~\cite{chen2017} to learn the distribution of travel times and distances between and within zones.
These distributions can be conditioned on time of day, day of the week, or season to more accurately model travel cost.
We found that for the New York Taxi Dataset, not all zones are directly connected (i.e., in a one-year period, no passengers made any trips between certain zones).
To model this sparse connectivity, we compute the least expected time path~\cite{miller-hooks2000} between unconnected zones and, calculate a distance and time distribution assuming this path of travel is used.
During simulation, when a vehicle moves between zones, these distributions are sampled to calculate required travel time and energy consumption as a function of a vehicle's energy efficiency and distance.

\subsection{Demand Model}
\begin{figure}
    \centering
    \subfigure[Vehicle State Diagram.]{
        \includegraphics[width=0.8\linewidth]{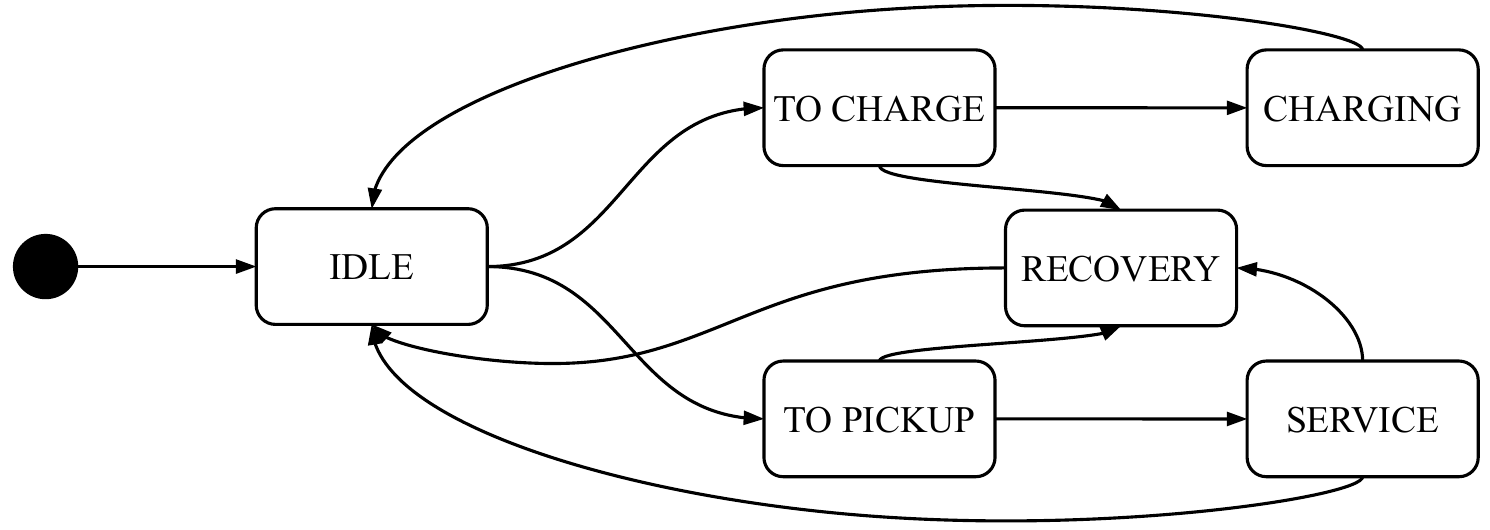}
        \label{fig:veh_sd}
    }
    \subfigure[Job State Diagram.]{
        \includegraphics[width=0.8\linewidth]{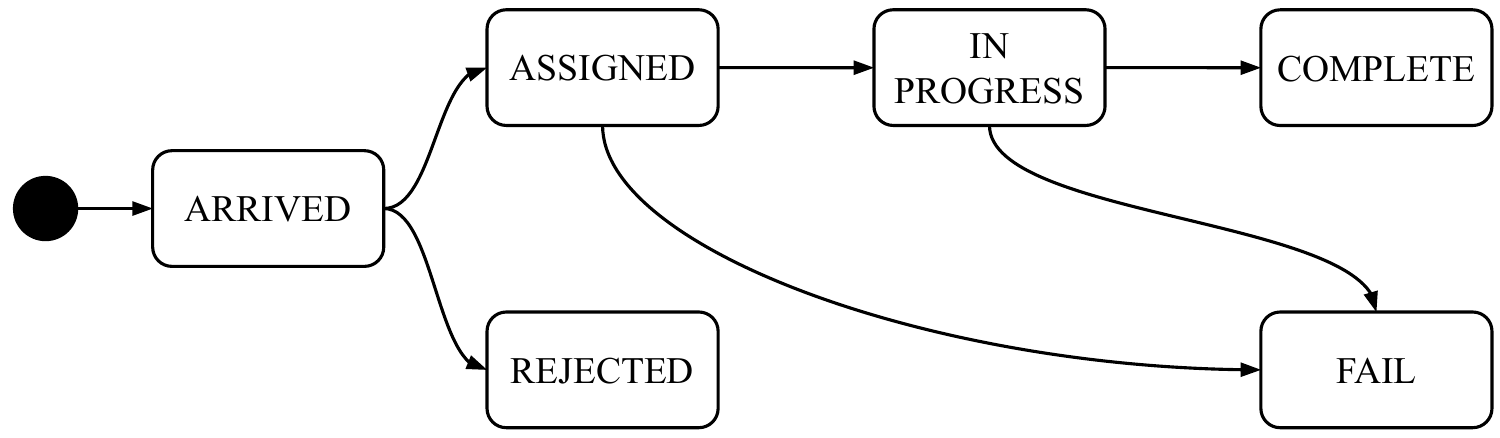}
        \label{fig:job_sd}
    }
    \caption{State diagrams of simulator entities}
\end{figure}

Jobs are simulated by iterating through entries in a real-world taxi dataset (e.g.~\cite{nyc-dataset}) in order of pick-up time.
In the simulator, each job $j$ is given as a tuple $\langle r_j, l_{pu,j}, l_{do,j}, \sigma_j, v_j \rangle$ where $r_j$ is the release time of the job, $l_{pu,j}$ is the pickup location, $l_{do,j}$ is the drop-off location, $\sigma_j$ is the job state, and $v_j$ is the vehicle currently assigned to the job.
The state of each job evolves according to Fig~\ref{fig:job_sd}.
Upon creation, the job enters the \textsc{arrived} state, after which it can be \textsc{rejected} or \textsc{assigned}.
Rejection occurs automatically if a job stays in the \textsc{arrived} state too long (1 hour in our experiments).
\textsc{assigned} means that a vehicle has been dispatched to service the demand and \textsc{in progress} indicates the passenger has been picked up.
The job completes after the vehicle reaches the drop-off location as indicated by the dataset.
A job can \textsc{fail} if the vehicle's SoC drops to $0$ at any point before completion.

\subsection{Grid and Charging Model}
The simulator considers a set of charging stations $\mathcal{S}$ where each $s\in\mathcal{S}$ is given by the tuple $\langle l_s, \bar{P}_s, \eta_s,  \mathcal{P}_s\rangle$.
$l_s$ is the zone in which the station is located, $\bar{P}_s$ is the maximum power that can be supplied by the station (which can be less than sum of the maximum power output of all ports), $\eta_s$ is the station's charging efficiency, and $\mathcal{P}_s$ is the set of ports.
Each $p\in\mathcal{P}_s$ is given by the tuple $\langle v_p, P_p(t),\bar{P}_p\rangle$ where $v_p$ is the vehicle assigned to charge at that port, $P_p(t)$ is the power delivered by that port on the interval $[t-1,t)$, and $\bar{P}_p$ is the port's maximum output power.

\subsection{Scheduler}
At every discrete time $t$, the scheduler receives $\mathcal{V}$, $\mathcal{S}$, and $\mathcal{J}$ (the set of all jobs released up to $t$) from the simulator.
A valid schedule consists of a set of instructions for each $v\in\mathcal{V}$.
Vehicles can either be assigned a job, a charger, or left to idle.
The simulator validates each received schedule and tracks the progress of vehicles toward their new assignments.
Old assignments can be overwritten by the scheduler at future time instances (e.g., in the case new ride demand arrives).

\section{Charge Scheduling with Reinforcement Learning}
We formulate the charge scheduling problem as an optimization problem with respect to the three indicators mentioned in the introduction: grid stress, battery degradation, and profitability over a finite time window $t_{min} \leq t < t_{max}$.
The maximization problem is illustrated in expression (\ref{eq:reward}).
The decision variable is $C_v(t)$, the charge rate allocated to any vehicle at a given time.  Note that if the scheduler allocates a charge rate to a vehicle not currently charging, the simulator enforces the travel time to the nearest station before charging begins.
The first term addresses profitability by attempting to maximize the number of jobs that have entered the \textsc{complete} state before $t_{max}$.
The second term addresses battery degradation by subtracting the total change in battery capacity of the fleet over the interval and multiplying weighting it by hyperparameter $\lambda$.
Finally, to address grid stress, we introduce a constraint where the total power that can be drawn by the entire fleet at any time $t$ on the interval must be less than hyperparameter $\delta$.
\begin{align*}
    \underset{C_v(t)}{\operatorname{max}} \quad & \sum_{j\in\mathcal{J}}{\mathbbm{1}_{\sigma_j=\textsc{complete}}} - \lambda\sum_{v\in\mathcal{V}}{\bar{Q}_v(t_{min}) - \bar{Q}_v(t_{max})}\\
    \text{s.t.}\quad &\sum_{s\in\mathcal{S}}{\sum_{p\in\mathcal{P}_s}{P_p(t)}}<\delta \quad \forall t_{min} \leq t < t_{max} \addtocounter{equation}{1}\tag{\theequation} \label{eq:reward}
\end{align*}
We use Lagrangian multipliers to transform the maximization problem in (\ref{eq:reward}) into the reward function for a reinforcement learning algorithm~\cite{gao2025}.
Since we have a continuous action space, we select proximal policy optimization reinforcement learning (PPO-RL)~\cite{xu2024} to train a policy neural network that outputs a charge schedule at every time $t$ given the simulator state.
Since we only address charge scheduling and not vehicle routing, we assume a fixed routing policy where vehicles requested to charge immediately stop their current task and begin traveling to the closest charger.
Likewise, vehicles which are not charging are dispatched to service the closest ride demand, if available.
For training, we set $\lambda=100$, $\delta=500$~kW, and train in Stable Baselines3~\cite{raffin2021} for 1e4 steps with episode lengths $168$ and $\Delta t=1$ hour.
The policy network has $2\times|\mathcal{V}|$ input neurons (SoH and SoC for each vehicle at $t$) and two intermediate layers of $64$ neurons each.  The output layer is also $2\times|\mathcal{V}|$ neurons with one output per vehicle giving a binary charge decision and the other giving the desired rate.  The neurons that determine whether to charge are bounded between $0$ and $1$ with a decision threshold at $0.5$, whereas the charge rate outputs are lower-bounded by $0$ and upper-bounded by a vehicle's maximum allowable charge rate.

\section{Experiments}
All code for our experiments has been made open source~\footnote{https://github.com/sccicitb/simulasi-pengisian-taksi-listrik}.
We compare our PPO-RL strategy with a baseline where fleet vehicles do not charge until they reach 20\% battery capacity and then charge at the maximum rate to 80\%~\cite{davies2019}.
Similar to PPO-RL, jobs are automatically assigned to the closest vehicle not currently charging, and vehicles are assigned to chargers based on their proximity.
Both policies were evaluated on a fleet of 50 taxis in New York City with 2 charging stations of 10 ports each and 50 kW max. charging power per port.  We used performance stats from the BYD e6~\cite{byd2021} to generate numbers for taxi battery capacity and efficiency.
We simulated this fleet for a period of 5 years to monitor the grid stress, battery health, and profitability under the two charging policies.
The parameters used for this experiment are detailed in Table~\ref{tab:exp}.
We place one charging station at Newark Airport (zone 1) and another in the Bronx (zone 208), outside of lower Manhattan, as it would be costly for a fleet operator to establish charging infrastructure in the downtown area.

\begin{table}
  \centering
  \caption{Experiment parameters. All vehicles in the fleet and all charging stations are identical except for locations.}
  \label{tab:exp}
  \begin{tabular}{p{0.4\linewidth} p{0.4\linewidth}}
    \hline
    \textbf{Parameter}&\textbf{Value}\\
    \hline
    $|\mathcal{V}|$ & 50 vehicles\\
    $\bar{Q}_{v}(0)$ & 71.7 kWh\\
    $\eta_v$ & 17.1 kWh per 100 km\\
    $l_s$ & $\{1, 208\}$\\
    $|\mathcal{P}_s|$ & 10 ports per station \\
    $\bar{P}_s$ & 500 kW \\
    $\eta_s$ & 0.9 (90\%)\\
    $\bar{P}_p$ & 50 kW \\
    $\Delta t$ & 1 hour\\
    $t_{max}$ & 5 years \\
  \hline
\end{tabular}
\end{table}

\subsection{Evaluation}
\begin{figure}
    \centering
    \includegraphics[width=1\linewidth]{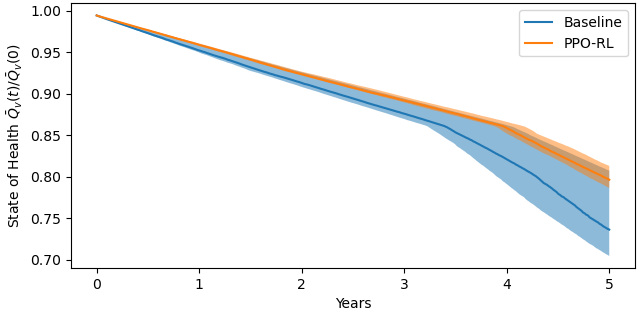}
    \caption{SoH across five years for simulated fleets charging with the baseline and PPO-RL charging policies.  The shaded area indicates the IQR across all vehicles in the fleet.}
    \label{fig:soh}
\end{figure}

\begin{figure}
    \centering
    \includegraphics[width=1\linewidth]{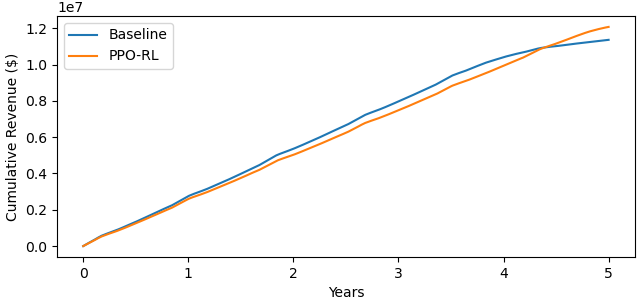}
    \caption{Cumulative revenue of a fleet charging under the baseline and the PPO-RL policies over 5 years.  PPO-RL maximizes revenue in the long run.}
    \label{fig:avail}
\end{figure}

Fig.~\ref{fig:soh} shows the SoH of both fleets over the 5 year period.
The shaded areas represent the interquartile range (IQR) of SoH across all vehicles and the solid line indicates the median SoH.
As expected, the SoH decays slower in the PPO-RL fleet.
We note that the decay is relatively linear for the first 3.5 years under the baseline (4 years under PPO-RL).
After this, the batteries begin a period of rapid degradation with almost all vehicles under 80\% SoH by the end of the $5^{th}$ year under the baselinem, whereas the upper $75^{th}$ percentile of vehicles is just approaching $80\%$ SoH under PPO-RL.  We mention $80\%$ SoH, because other works have suggested that vehicles should be taken out of service when SoH falls below this value~\cite{wan2024}.
In Fig.~\ref{fig:avail} we calculate cumulative revenue across the 5 years (based on fares reported in the New York Taxi Dataset) under both policies when vehicles crossing the $80\%$ SoH threshold are removed from the simulation.
Due to the decrease in suitable vehicles, the revenue returned by the baseline fleet tapers off after 4 years, whereas under PPO-RL it continues to accumulate at roughly the same rate.
This means the baseline charging policy is more profitable for the first $4.5$ years, but beyond that, PPO-RL is better.

\begin{figure}
    \centering
    \includegraphics[width=1\linewidth]{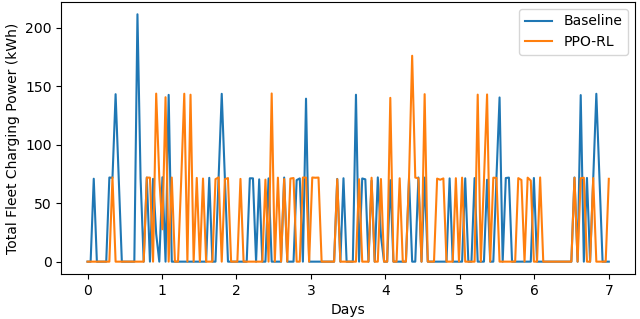}
    \caption{Total fleet charging power drawn from the grid over a typical week under the baseline and PPO-RL charging policies.  The baseline has a higher peak power when multiple vehicles' charging times align.}
    \label{fig:ch_power}
\end{figure}

Finally, we evaluate grid stress by plotting the hourly total charging power consumed by the fleet over a typical week in Fig.~\ref{fig:ch_power}.
Under the baseline policy, individual vehicles coming to charge cause large spike in electricity demand, and these can exceed 200~kW when multiple charging sessions overlap.
With PPO-RL, the charging sessions are spread out over time, leading to less stress on the grid.

\subsection{Cross-Environment Evaluation}
\begin{figure}
    \centering
    \includegraphics[width=1\linewidth]{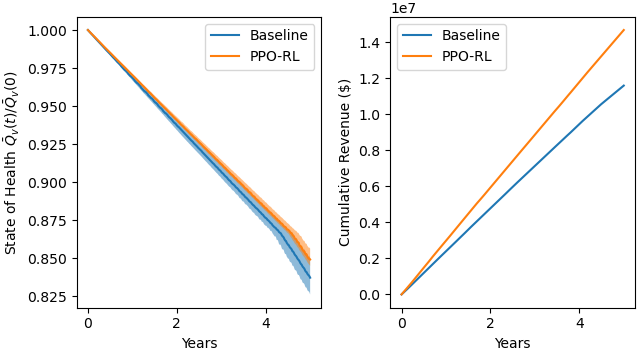}
    \caption{Evaluation of a fleet trained on New York demand data but operating in Chicago.  Left: SoH across five years; the shaded area indicates the IQR across all vehicles in the fleet.  Right: Cumulative revenue of a fleet charging under the baseline and the PPO-RL policies.}
    \label{fig:chicago}
\end{figure}

\begin{figure}
    \centering
    \includegraphics[width=1\linewidth]{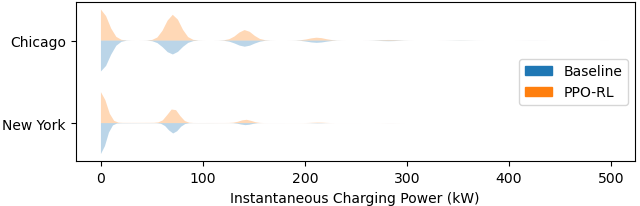}
    \caption{Distribution of instantaneous electric power consumption under the baseline and PPO-RL policies across the entire the entire five year life of the fleet in the New York and Chicago regions.}
    \label{fig:dist}
\end{figure}

We also want to check if our learned policy generalizes well to other cities.  The  City of Chicago Taxi Rides Dataset~\cite{chicago-dataset} structures ridership data similar to the New York Taxi Dataset, so we compare the performance of the policy trained in New York and applied in Chicago to the baseline policy.  We consider the same number of vehicles and move the charging stations to O'Hare Airport and Hegewisch to mimic the setup in New York.
The results for state of health and cumulative revenue are plotted in Fig.~\ref{fig:chicago}.  First, we observe that for both policies the mean state of health after five years is higher than in New York City (85\% compared to 80\% and 75\% for PPO-RL and the baseline respectively).  We believe this is due to the lower ride demand in Chicago, which means each vehicle serves fewer rides over its lifespan.  Once again, the PPO-RL outperforms the baseline in terms of mean battery health at the end of the five-year period.  We also observe that the PPO-RL model now achieves a higher cumulative profit across the entire life of the fleet.  We believe that with fewer available rides, PPO-RL is more likely to have some partially charged vehicles ready to service demand at any give time.  We note that the discrepancy in results between New York and Chicago indicates that more work is needed regarding the generalization of learned policies. Finally, in Fig.~\ref{fig:dist} we compare the distribution of grid power required by PPO-RL and the baseline; both perform comparably with clear distribution modes when multiple vehicles charge simultaneously.

\section{Conclusion}
Charge scheduling in EV taxi fleets is a complex problem.
Deciding where and when EVs should charge impacts the fleet's ability to service demand, the future health of the fleet's batteries, and imposes a certain amount of stress on the grid.
We proposed combining these aspects: ride demand, grid conditions, charging infrastructure, and battery degradation, into one simulator to better evaluate and develop charge scheduling policies.
To demonstrate the utility of this simulator, we assumed a simple routing policy for vehicle-to-ride and vehicle-to-charger allocation and used reinforcement learning to train a charging policy that balanced profitability, battery health, and grid stress.
While our training method was simple, the improvement over a baseline policy typically employed in EV taxi fleets showed that charge scheduling has a great potential to improve fleet operations.

We plan to develop this simulator further to better model technologies like vehicle-to-grid, and demand response.
Additionally, we would like to develop more complex behavior models for passengers and drivers (e.g., a charge schedule may need to be adapted to individual driving styles) as well as modeling charging station and vehicle downtime due to maintenance.
We encourage community contributions in these areas and others which may be of academic or industrial interest.
We would also like to develop robust charging policies that are explainable and reliable in the face of unpredictable environmental conditions.
This simulator can help with such what-if analyses and we hope that in the future it can serve as the basis of a digital twin for real-world electric taxi fleets.

%\section*{Acknowledgment}

%The preferred spelling of the word ``acknowledgment'' in America is without 
%an ``e'' after the ``g''. Avoid the stilted expression ``one of us (R. B. 
%G.) thanks $\ldots$''. Instead, try ``R. B. G. thanks$\ldots$''. Put sponsor 
%acknowledgments in the unnumbered footnote on the first page.

%\section*{References}

\bibliographystyle{IEEETran}
\bibliography{references}

\end{document}